\documentclass[ aip,
 amsmath,amssymb,floatfix,
 reprint,%
]{revtex4-1}

\usepackage{graphicx}
\usepackage{dcolumn}
\usepackage{bm}

\usepackage[utf8]{inputenc}
\usepackage[T1]{fontenc}
\usepackage{mathptmx}
\usepackage{etoolbox}
\usepackage{placeins}
\usepackage{comment}

\usepackage{siunitx}
\usepackage[version=4]{mhchem}

\makeatletter
\def\@email#1#2{%
 \endgroup
 \patchcmd{\titleblock@produce}
  {\frontmatter@RRAPformat}
  {\frontmatter@RRAPformat{\produce@RRAP{*#1\href{mailto:#2}{#2}}}\frontmatter@RRAPformat}
  {}{}
}%
\makeatother

\begin{document}

\title{Correlated domain and crystallographic orientation mapping in uniaxial ferroelectric polycrystals by interferometric vector piezoresponse force microscopy} 

\author{R. Dragland}
\affiliation{
Department of Materials Science and Technology, NTNU Norwegian University of Science and Technology, Trondheim, Norway
}%
\author{J. Schulthei\ss{}}
\email[]{jan.schultheiss@ntnu.no}
\affiliation{
Department of Materials Science and Technology, NTNU Norwegian University of Science and Technology, Trondheim, Norway
}%
\author{I. N. Ushakov}
\affiliation{
Department of Materials Science and Technology, NTNU Norwegian University of Science and Technology, Trondheim, Norway
}%
\author{R. Proksch}
\affiliation{ Asylum Research an Oxford Instruments Company, Santa Barbara, California, USA
}%
\author{D. Meier}
\email[]{dennis.meier@uni-due.de}
\affiliation{
Department of Materials Science and Technology, NTNU Norwegian University of Science and Technology, Trondheim, Norway
}%
\affiliation{
Faculty of Physics and Center for Nanointegration Duisburg-Essen (CENIDE), University of Duisburg-Essen, Duisburg, Germany
}%
\affiliation{
Research Center Future Energy Materials and Systems, Research Alliance Ruhr, Bochum, Germany
}

\begin{abstract}
Ongoing advances in scanning probe microscopy techniques are continually expanding the possibilities for nanoscale characterization and correlated studies of functional materials. Here, we demonstrate how a recent extension of piezoresponse force microscopy (PFM), known as interferometric vector PFM, can be utilized for simultaneously mapping the local crystallographic orientations and the domain structure of distributed grains in uniaxial ferroelectric polycrystals. By shifting the laser beam position on the cantilever, direction-dependent piezoresponse signals are acquired analogous to classical vector PFM, but without the need to rotate the sample. Using polycrystalline \ce{ErMnO3} as a model system, we demonstrate that the reconstructed piezoresponse vectors correlate one-to-one with the crystallographic orientations of the micrometer-sized grains, carrying grain-orientation and domain-related information. We establish a versatile approach for rapid, multimodal characterization of polycrystalline uniaxial ferroelectrics, enabling automated, high-throughput reconstruction of polarization and grain orientations with nanoscale precision.
\end{abstract}

\maketitle 

\newcommand{\EMO}[1]{ErMnO$_3$#1}

\noindent
Piezoresponse force microscopy (PFM) has evolved into a mainstream technique for local characterization of ferroelectric and related materials that display a piezoelectric effect\cite{soergel_piezoresponse_2011, gruverman_piezoresponse_2006, gruverman_piezoresponse_2019}.
PFM is essential for analyzing domain and domain wall structures\cite{gruverman1996nanoscale}, including their electromechanical and dynamical responses \cite{alexe1999patterning}, and interactions with defects, such as topological solitons~\cite{ivry2010flux}, precipitates~\cite{zhao_precipitation_2021}, or grain boundaries~\cite{marincel_influence_2014}. This non-exhaustive list of examples for the application of PFM reflects its versatility, and the methodology is continuously evolving, driving the development of more sophisticated and powerful measurement modes: Switching spectroscopy PFM (SS-PFM) enables local mapping of ferroelectric  hysteresis\cite{jesse_switching_2006}, stroboscopic PFM (S-PFM) provides access to nanosecond-scale domain dynamics in ferroelectric devices \cite{gruverman_direct_2005}, and dual AC resonance tracking PFM (DART-PFM)\cite{gannepalli_mapping_2011} and band-excitation PFM (BE-PFM)\cite{jesse_band_2011} allow for imaging of domain structures in materials with weak piezoelectric coefficients and spatially varying resonance frequencies.

To obtain complete information about the polarization orientation, vector piezoresponse force microscopy (vPFM) is a well-established and widely applied technique. By sequentially rotating and scanning the sample surface, vPFM enables visualization of the three-dimensional polarization vector \cite{kalinin_vector_2006}. 
Recently, it has been demonstrated that equivalent information can also be obtained by sequentially shifting the beam position on the cantilever instead of rotating the sample, a technique called interferometric vPFM~\cite{proksch_3d_2025}. 
In a first proof-of-concept study, domain orientations were successfully determined in a bismuth ferrite thin film. With its suitability for automation, interferometric vPFM provides a promising pathway toward high-throughput nanoscale characterization, which we further explore and expand in this work.

Here, we apply interferometric vPFM to gain correlated insights into the microstructure and domain distribution in uniaxial ferroelectric polycrystals with randomly oriented grains. By mapping the piezoresponse of an area with several grains using four different beam positions on the cantilever, we derive the direction-dependent piezoresponse and compare it with the crystallographic orientation of each grain measured by electron backscatter diffraction (EBSD) \cite{humphreys_review_2001}. The comparison shows that in addition to resolving the ferroelectric \SI{180}{\degree} domains within the grains, the direction-dependent piezoresponse exhibits a one-to-one correlation with EBSD data, providing information about the orientation of the individual grains, their statistical distribution, and texturing.

\begin{figure}
    \centering
    \includegraphics[width=\linewidth]{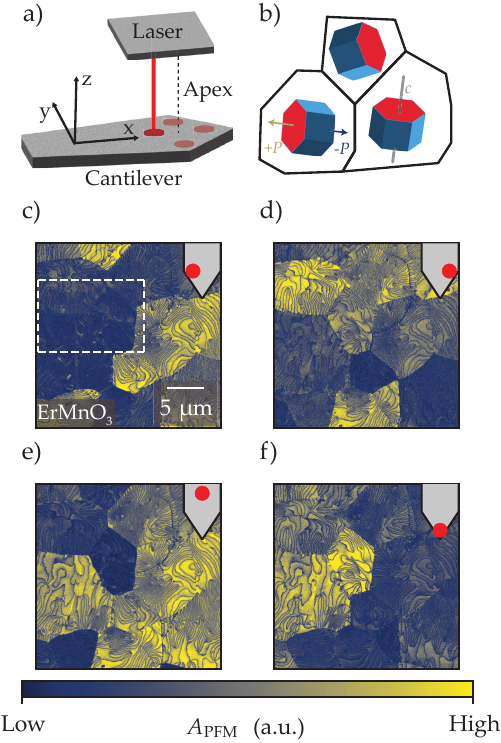}
    \caption{ 
    Shown in a) is a schematic drawing of the experimental setup for interferometric vPFM. Scans are performed with the laser beam (red line) sequentially positioned at each of the red dots. The position used for conventional PFM, with the beam placed directly above the AFM tip apex, is shown by a dashed line. The coordinate system is defined with $x$ along and $z$ normal to the cantilever. A schematic representation of the polycrystalline \EMO{} model system is shown in b). The hexagons represent the crystallographic orientation of the unit cells within a grain, with the polarization, $\pm P$, being parallel to the crystallographic $c$-axis. The acquired and background corrected piezoresponse amplitude $A_{\mathrm{PFM}}$ data (Supplementary Section II) shown in c–f) are recorded with the laser beam positioned at different locations along the cantilever, as indicated in the inset. The white dashed box in c) marks the area discussed in Figure \ref{fig:2fig_pfm_feature}. }
    \label{fig:1fig_raw}
\end{figure}

The basic concept of interferometric vPFM, as introduced in ref.\ \onlinecite{proksch_3d_2025}, is presented in Figure \ref{fig:1fig_raw} a). When performing conventional PFM, the laser is typically positioned at the projection of the tip apex onto the backside of the cantilever, schematically visualized by a dotted line. In the case of interferometric vPFM, the laser beam is deliberately displaced to obtain directional piezoresponse contrast. To describe this directional contrast, we define a right-handed coordinate system with $x$ along the cantilever length, $y$ along its width, and $z$ normal to its surface. Polycrystalline \EMO{} is used as a model system to test the capability of interferometric vPFM to simultaneously map grain orientations and ferroelectric domain structures. \EMO{} is a uniaxial ferroelectric with geometrically driven spontaneous polarization ($T_{\mathrm{C}} =$ \SI{1150}{\celsius}, $P$ = \SI{5.6}{\micro\coulomb\per\square\cm}), forming intriguing patterns of \SI{180}{\degree} domains with sixfold meeting points as explained elsewhere~\cite{jungk_electrostatic_2010,choi_insulating_2010,van_aken_origin_2004}. The polycrystals, synthesized via a solid-state reaction route, consist of individual crystallites with randomly oriented $c$-axes~\cite{schultheis_confinementdriven_2022}, each exhibiting a spontaneous polarization $\pm P$ parallel to the $c$-axis, as schematically illustrated by the hexagons in Figure \ref{fig:1fig_raw} b). Details on the synthesis are provided in Supplementary Section I and ref.~\onlinecite{schultheis_confinementdriven_2022}.

We begin our analysis with the interferometric vPFM data set shown in Figure \ref{fig:1fig_raw} c–f), obtained by scanning the same region with different beam positions on the cantilever and correcting the background offset as described in Supplementary Section II. A Cypher system (Oxford Instrument, Santa Barbara, CA, USA) is used for this data acquisition, with more details on the experimental conditions provided in Supplementary Section III. The corresponding cantilever schematics, shown in the top right of each scan, indicate the beam positions shifted in the $-y$-, $+y$-, $-x$-, and $+x$-directions. All scans exhibit pronounced intergranular contrast, indicating the ability of interferometric vPFM to distinguish between different grains. Importantly for our analysis, this grain-selective contrast is related to the laser position on the cantilever, which we will explore in more detail in the following.

\begin{figure}
    \centering
    \includegraphics[width=\linewidth]{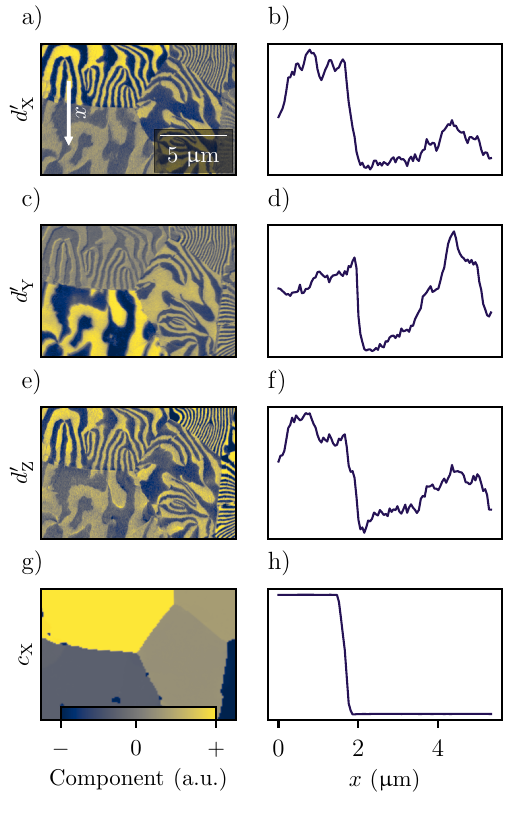}
    \caption{Direction-dependent piezoresponse maps, $d'_\mathrm{X}$, $d'_\mathrm{Y}$, and $d'_\mathrm{Z}$, calculated from the raw data highlighted by the dashed white box in Figure \ref{fig:1fig_raw}, are shown in a), c), and e), respectively. Cross sections along the arrow in a) are extracted in b), d), and f) and reveal pronounced variations in piezoresponse. The crystallographic $c_{\mathrm{X}}$ component, measured via EBSD at the same position, is shown in g), with the line-profile presented in h). A clear correlation between the $d'_\mathrm{X}$ data, obtained from interferometric vPFM in b), and $c_\mathrm{X}$, measured via EBSD in h), can be observed.}
    \label{fig:2fig_pfm_feature}
\end{figure}

To derive the direction-dependent piezoresponses, $d'_\mathrm{X}$, $d'_\mathrm{Y}$, $d'_\mathrm{Z}$ we apply the transformation presented in ref. \onlinecite{proksch_3d_2025}, combining the different vPFM signals in Fig \ref{fig:1fig_raw} c-f). Note that for simplicity, we write $d'_{\mathrm{j}}$ instead of $d_{\mathrm{eff,j}}$ in this work. 
The calculated $d'_\mathrm{X}$, $d'_\mathrm{Y}$, and $d'_\mathrm{Z}$ are shown in Figure \ref{fig:2fig_pfm_feature} a), c), and e), respectively, for the selected region marked by a dashed box in Figure \ref{fig:1fig_raw} c). All images clearly show vortex- and stripe-like domain structures within the grains, revealing the characteristic domain patterns of polycrystalline hexagonal manganites \cite{schultheis_confinementdriven_2022,dragland_relation_2025,wolk_coexistence_2024}. Furthermore, the grain-dependent contrast variations are preserved by the transformation, corroborating that the grains have different direction-dependent piezoresponses. Cross sections across two grains, shown in Figure \ref{fig:2fig_pfm_feature} b), d), and f), highlight the abrupt change in contrast at one of the grain boundaries. To confirm that these contrast changes are indeed grain-orientation dependent, we correlate the interferometric vPFM data with EBSD data obtained at the same position. The comparison of the measurements reveals that the $c_\mathrm{X}$ component of the EBSD data, displayed in Figure \ref{fig:2fig_pfm_feature} g-h), correlates with the interferometric vPFM data with a pronounced contrast change at the grain boundary.

\begin{figure}
    \centering
    \includegraphics{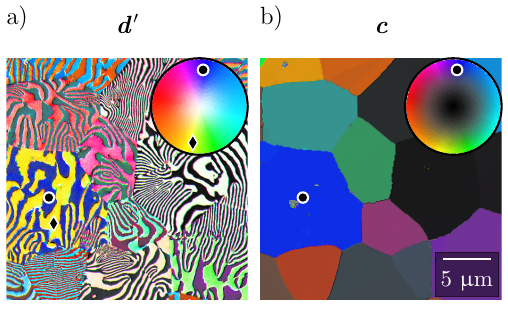}
    \caption{    
    The direction-dependent piezoresponse vector, $\boldsymbol{d'}$, is visualized for the entire scan area in a). The color-code using hue, represents the in-plane piezoreponse, while the brightness and saturation quantify the component directed out-of-plane, $d'_{\mathrm{Z}}$. A map visualizing the crystallographic $\boldsymbol{c}$-axis, obtained via EBSD, is included in b). The ability to resolve the \SI{180}{\degree} domain structure is demonstrated for the blue/yellow domain within one grain in a), indicated by the square and triangle on opposite sites of the color wheel. In this representation, the $d'_{\mathrm{Z}}$ is scaled to match the range of the in-plane components, as it is typically about twice as large in magnitude.\cite{proksch_3d_2025}}
    \label{fig:3fig_comparison_2D}

\end{figure}

To demonstrate the general validity of this correlation, we combine the direction-dependent vPFM signals into a single data set for the entire measured area, shown in Figure \ref{fig:3fig_comparison_2D} a). The in-plane angle of the piezoresponse is color-coded using hue, whereas the out-of-plane component is represented by saturation and brightness. This three-dimensional map allows for distinguishing individual grains and visualizes both their ferroelectric domain structure and the polarization orientations. Grains displaying vibrant colors are predominantly in-plane, whereas grains with black-and-white colors are oriented out-of-plane. 
An example demonstrating the resolution of \SI{180}{\degree} domain separation is shown for the grains highlighted in Figure \ref{fig:3fig_comparison_2D} a). The expected \SI{180}{\degree} domain shift is indicated on the color wheel by the blue and yellow regions, marked with a square and a triangle, respectively. For comparison, an EBSD map recorded at the same position is shown in Figure \ref{fig:3fig_comparison_2D} b), confirming that the grain separation obtained from the interferometric vPFM data accurately reflects the underlying crystallography.

\begin{figure*}
    \centering
    \includegraphics{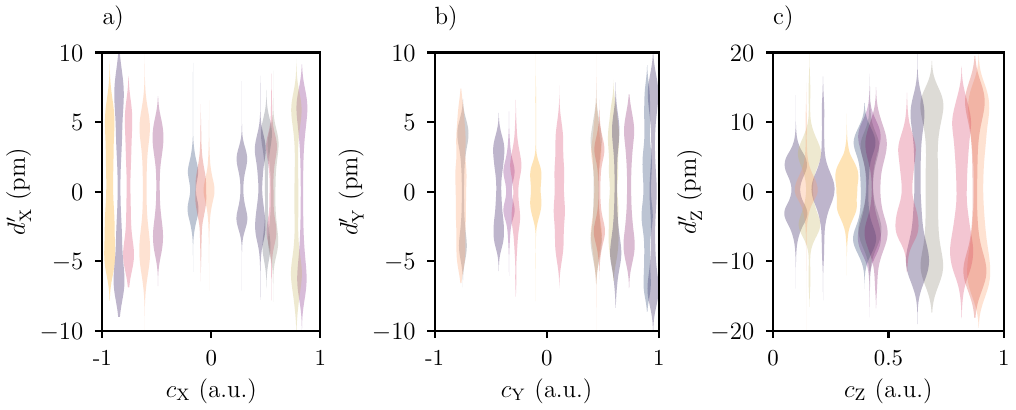}
    \caption{    
     The direction-dependent distributions of the piezoresponse are displayed as a function of the corresponding EBSD components for the \num{15} mapped grains. The sampled piezoresponses reflect bimodal distributions that broaden with increasing EBSD-component magnitude, reflecting the antiparallel alignment of ferroelectric domains within one grain.}
    \label{fig:5fig_correlation}

\end{figure*}

\begin{figure}
    \centering
    \includegraphics[width=\linewidth]{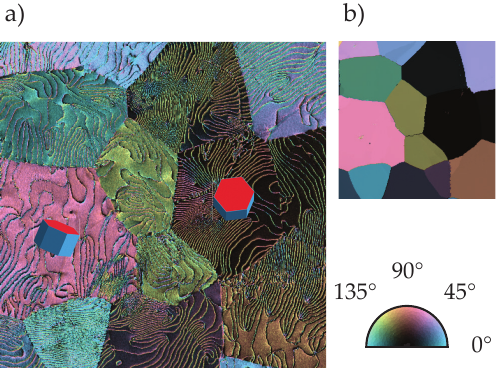}
    \caption{
    From the three-dimensional piezoresponse data in Figure \ref{fig:3fig_comparison_2D}, the crystallographic $c$-axis direction is calculated and presented in a), with two hexagons highlighting a predominantly in-plane and a predominantly out-of-plane grain. The reference EBSD map, presented in Figure \ref{fig:3fig_comparison_2D} b), is displayed in b) for comparison. The in-plane component is encoded via hue (color), while the out-of-plane component is encoded by the brightness. Higher out-of-plane component corresponds to lower brightness.}
    \label{fig:4fig_comparison_3D}

\end{figure}

A quantitative analysis of the correlation between grain orientations derived from interferometric vPFM and EBSD is presented in Figure \ref{fig:5fig_correlation}. The direction-dependent piezoresponses obtained from interferometric vPFM are compared with the corresponding EBSD signal components for \num{15} grains. A clear trend between interferometric vPFM and EBSD can be observed. Specifically, grains that have a preferential crystallographic orientation along a given direction also exhibit a higher absolute piezoelectric response in that direction. For example, in Figure \ref{fig:5fig_correlation} a), grains with negligible $c_{\mathrm{X}}$ component show a small, unimodal piezoresponse distribution, whereas highly oriented grains display a clearly bimodal distribution with piezoresponse values up to an order of magnitude larger. Figure \ref{fig:5fig_correlation} b) and c) show that a similar trend is observed for the $d'_\mathrm{Y}$ and $d'_\mathrm{Z}$ piezoresponses, corroborating that interferometric vPFM can reliably quantify all components of the grain orientations. The grain orientations obtained from the three-dimensional interferometric vPFM map are finally visualized in Figure \ref{fig:4fig_comparison_3D} a), with two exemplary grains, one primarily in-plane and the other primarily out-of-plane, highlighted by hexagons. For comparison, the corresponding EBSD map is shown in Figure \ref{fig:4fig_comparison_3D} b).

In conclusion, using uniaxial ferroelectric \EMO{} polycrystals as the model system, we have demonstrated that interferometric vPFM can simultaneously map crystallographic orientation and ferroelectric domain structure. By comparing the direction-dependent piezoresponses to EBSD signals, we confirm that four sequential beam-shifted scans enable spatially resolved measurements of grain orientations and ferroelectric domain structure without the need to rotate the sample. We find a one-to-one correlation between the grain orientations obtained from interferometric vPFM and EBSD data. Going beyond previous studies that demonstrated orientational mapping in single-crystalline thin films \cite{proksch_3d_2025}, we extend interferometric vPFM towards correlated nanoscale studies in polycrystalline materials.

Our method can be readily transferred to other polycrystalline uniaxial ferroelectrics, including isostructural hexagonal oxides~\cite{nordlander_epitaxy_2022}, lithium niobate~\cite{miller_temperature_1966} or strontium-barium niobate~\cite{lukasiewicz_strontiumbarium_2008} where polycrystallinity and associated texturing plays a key role for the physical properties. Moreover, it can be applied to combined domain imaging and grain-orientation determination in more complex multi-axial polycrystalline ferroelectrics, relevant for sensors, actuators, high-power devices, and biomedical applications~\cite{koruza_requirements_2018, zhang_piezoelectric_2005, thu_potential_2025}. Leveraging the simplicity of the method compared to classical vector PFM, which requires physical rotation of the sample, and the feasibility of automating sequential beam-shifted scans and subsequent data analysis, our approach aligns with automated PFM characterization efforts~\cite{kalinin_automated_2021} and can be readily implemented in high-throughput ceramic processing workflows~\cite{eckstein_feasibility_2024}.

\noindent
\textbf{Acknowledgements}\\
R.D., J.S. and D.M. acknowledge funding from the European Research Council (ERC) under the European Union’s
Horizon 2020 Research and Innovation Program (Grant Agreement No. 863691). D.M. thanks NTNU for support through the Onsager Fellowship Program, the outstanding Academic Fellow Program.

\noindent

\subsection*{References}
\bibliographystyle{MSP}
\bibliography{IvPFM_EMO}

\end{document}



\title{Correlated domain and crystallographic orientation mapping in uniaxial ferroelectric polycrystals by interferometric vector piezoresponse force microscopy} 

\author{R. Dragland}
\affiliation{
Department of Materials Science and Technology, NTNU Norwegian University of Science and Technology, Trondheim, Norway
}%
\author{J. Schulthei\ss{}}
\email[]{jan.schultheiss@ntnu.no}
\affiliation{
Department of Materials Science and Technology, NTNU Norwegian University of Science and Technology, Trondheim, Norway
}%
\author{I. N. Ushakov}
\affiliation{
Department of Materials Science and Technology, NTNU Norwegian University of Science and Technology, Trondheim, Norway
}%
\author{R. Proksch}
\affiliation{ Asylum Research an Oxford Instruments Company, Santa Barbara, California, USA
}%
\author{D. Meier}
\email[]{dennis.meier@uni-due.de}
\affiliation{
Department of Materials Science and Technology, NTNU Norwegian University of Science and Technology, Trondheim, Norway
}%
\affiliation{
Faculty of Physics and Center for Nanointegration Duisburg-Essen (CENIDE), University of Duisburg-Essen, Duisburg, Germany
}%
\affiliation{
Research Center Future Energy Materials and Systems, Research Alliance Ruhr, Bochum, Germany
}


\maketitle

\newcommand{\EMO}[1]{ErMnO$_3$#1}

\section{Synthesis}
\noindent
Polycrystalline \EMO{} was synthesized via a solid-state reaction from Er$_2$O$_3$ (\SI{99.9}{\percent}) and Mn$_2$O$_3$ (\SI{99.0}{\percent}, Alfa Aesar). Stoichiometric powders were ball-milled for \SI{24}{\hour} at \SI{205} rpm, heat-treated at \SI{1100}{\celsius} for \SI{12}{\hour}, pressed, and densified at \SI{1350}{\celsius} for \SI{12}{\hour}. The polycrystalline sample surface was prepared for PFM by lapping using a \SI{9}{\micro\meter} Al$_2$O$_3$ water suspension (Logitech, Glasgow, Scotland), followed by polishing utilizing a silica slurry (SF1 Polishing fluid, Logitech). A detailed synthesis description is given in ref.~\onlinecite{schultheis_confinementdriven_2022}.

\section{Background subtraction of PFM data}
\noindent
In the following is an elaborate description of the performed post-processing of the obtained PFM images to account for noise with the same frequency as the lock-in reference frequency. 
When AC-bias is applied to a conductive tip in contact with a piezoelectric material, the material will start to locally vibrate due to converse piezoelectric effect. The vibrational component of frequency $\omega$ at a surface point/tip position (\textit{x},\textit{y}) can be described as the real part of
\begin{equation}
    P_{\omega}(x,y,t) = A_{\omega}e^{i(\phi_{\omega}(x,y)+\omega t)},
\end{equation}
\noindent
where $A_\omega(x,y) \in \mathbb{R}$ is the piezoelectric amplitude, $\phi_\omega(x,y) \in \mathbb{R}$ is the piezoelectric phase, and $\omega$ is the frequency read by the lock-in amplifier (typically set equal to the frequency of applied AC-bias). In the following, we drop the subscripts $\omega$.

\begin{figure}
    \centering
    \includegraphics[width=\linewidth]{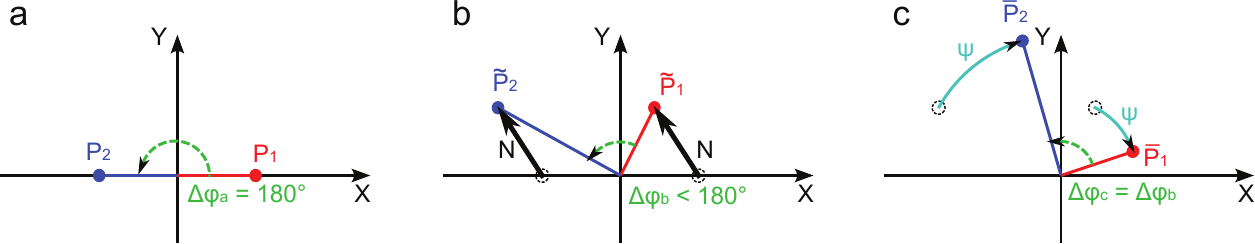}
    \caption{Signal warping in PFM. a) Given piezoresponse phasors $P_1$ and $P_2$ at two different positions, b) the noise $N(t)$ displaces the phasors, changing their amplitudes and phase difference $\Delta\phi$. c) Leading/lagging currents equally rotate the phasors without changing their amplitudes and phase difference. However, the absolute phase values change.}
    \label{fig:s1}
\end{figure}

The measured piezoresponse will differ from the physically expected one, as our experimental setup will inevitably contain noise with Fourier component of frequency $\omega$ (coming from electronics, the environment or whatever external source)\cite{jungk_quantitative_2006}. We assume that the noise amplitude and phase are time-independent on the timescales of the measurement:

\begin{equation}
    N(t) = A_{\mathrm{N}} e^{i(\phi_{\mathrm{N}} +\omega t)}.
\end{equation}
\noindent
This transforms the measured piezoresponse to:

\begin{equation}
    \tilde{P}(x,y,t) = P(x,y,t) + N(t) = e^{i(\omega t)}[A(x,y) e^{i\phi_{\omega}(x,y)} + A_{\mathrm{N}} e^{i\phi_{\mathrm{N}}} ].
\end{equation}
\noindent
In addition, there will be an overall phase offset $\psi$ due to electronic delay \cite{soergel_piezoresponse_2011}:

\begin{equation}
    \bar{P}(x,y,t) = \tilde{P}(x,y,t)e^{-i\psi}.
\end{equation}
\noindent
The resultant effect on the measured phase and amplitude can be visualized in a phasor diagram as illustrated in Figure \ref{fig:s1}.

If we, based on symmetry or other physical arguments, can find constraints for the amplitudes and phases in different regions of the scanned surface, we could reverse engineer from the situation in Figure \ref{fig:s1} c) to the situation in Figure \ref{fig:s1} a) by either actively applying phase and noise offsets during the scan (with a lock-in amplifier), or passively introduce the offsets when post-processing the data. 
In \EMO{,} within one grain, the two ferroelectric domain states are expected (by symmetry) to give piezoresponses with same amplitudes, and a \SI{180}{\degree} phase difference\cite{kumagai_structural_2013}. Having two PFM images $[A_{\mathrm{ij}}]$ and $[\phi_{\mathrm{ij}}]$ representing respectively amplitude and phase at discrete pixels (i,j), we used following algorithm for post-processing:

\begin{enumerate}
    \item Pick two points (or two sets of points) (k,l) and (m,n), representing two different domain states within the same grain. Find $\psi$ such that $A_{\mathrm{kl}}\sin({\phi_{\mathrm{kl}} - \psi}) = A_{\mathrm{mn}}\sin({\phi_{\mathrm{mn}} - \psi})$. Then correct the phase: \newline
    $[\hat\phi_{\mathrm{ij}}] = [\phi_{\mathrm{ij}}] - \psi$. 
    \item Convert polar data to Cartesian data and add offsets $N_{\mathrm{X}}$ and $N_{\mathrm{Y}}$: \newline
    $[X_{\mathrm{ij}}] = [N_{\mathrm{X}}] + [A_{\mathrm{ij}}]\cos[\hat\phi_{\mathrm{ij}}] $, \newline
    $[Y_{\mathrm{ij}}] = [N_{\mathrm{Y}}] + [A_{\mathrm{ij}}]\sin[\hat\phi_{\mathrm{ij}}] $.
    \item Convert the resulting Cartesian data to polar data: \newline
    $[A_{\mathrm{ij}}] = \sqrt{ [X_{\mathrm{ij}}]^2 + [Y_{\mathrm{ij}}]^2 }$, \newline
    $[\phi_{\mathrm{ij}}] = \atantwo({[Y_{\mathrm{ij}}],[X_{\mathrm{ij}}]) }$.
    \item Repeat steps \num{2} and \num{3} for new values of $N_{\mathrm{X}}$ and $N_{\mathrm{Y}}$ until convergence has been reached. 
    \item Repeat steps \num{1}-\num{5} for different scans, consistently choosing same domains in step \num{1} and avoiding contrast inversion.

\end{enumerate}

\begin{figure}
    \centering
    \includegraphics[width=0.69\linewidth]{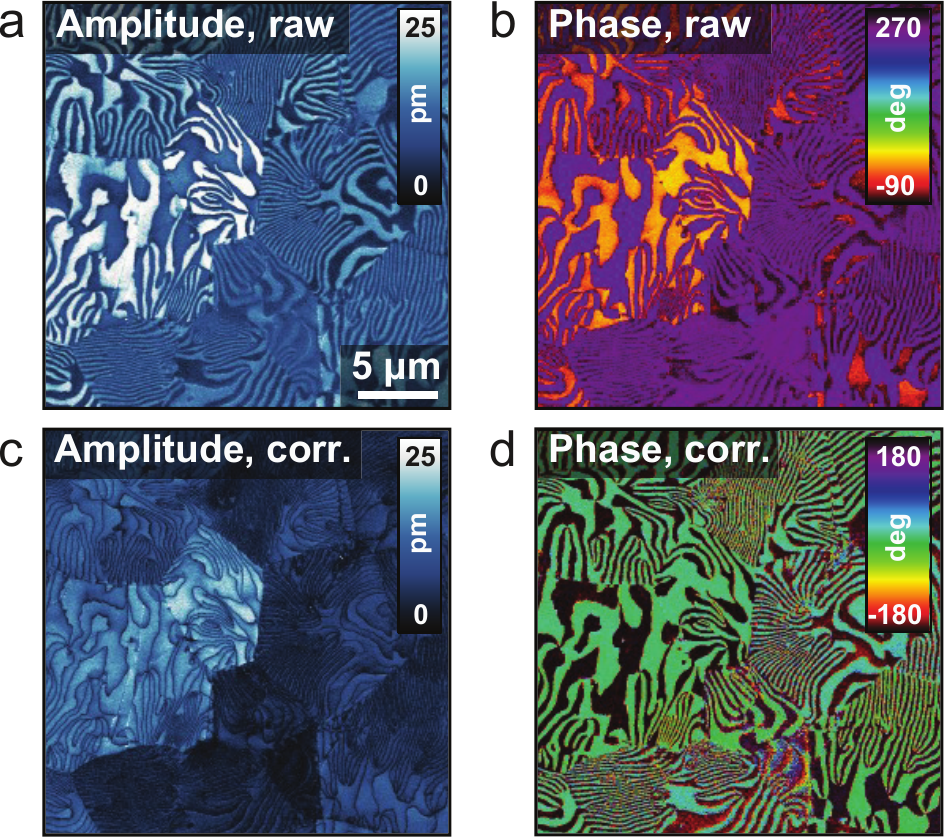}
    \caption{PFM correction. a) Amplitude image of a grain structure in \EMO{} (raw data). b) The phase for the same scan as in a). Note that the phase difference is less than \SI{180}{\degree} within each grain and varies from grain to grain. c) Corrected amplitude based on the post-processing algorithm. Note that the grains have different amplitudes, depending on the orientation, but the domains within the same grain have the same amplitude. Domain walls exhibit characteristic freezing. d) Corrected phase based on the post-processing algorithm. Note that different domains have \SI{180}{\degree} difference, with minimal variation between the grains (with some noise where amplitude approaches 0). }
    \label{fig:s2}
\end{figure}

Figure \ref{fig:s2} shows a corrected image obtained with the post-processing algorithm. The resulting image is in excellent agreement with what is expected physically – within each grain, the two domain states have the same amplitudes and are \SI{180}{\degree} out of phase. The phase does not vary from grain to grain. Note that the amplitude freezes at the domain walls, which is an expected effect \cite{kumagai_structural_2013}.

\section{Instrument settings}
\noindent
PFM data were acquired using a Cypher system (Oxford Instruments, Santa Barbara, CA, USA) equipped with an interferometric displacement sensor (IDS) and Adama AD-2.8-AS boron-doped diamond probes (Oxford Instruments, Santa Barbara, CA, USA) 
under an applied voltage of \SI{3}{\volt} and a frequency of \SI{55}{\kilo\hertz}.

EBSD data were collected using a Symmetry S3 detector (Oxford Instruments, High Wycombe, UK) on a Hitachi SU5000 thermally assisted FEG-SEM. 
\clearpage
\bibliographystyle{MSP}
\bibliography{IvPFM_EMO}